\documentclass[twocolumn,aps,showpacs,floatfix,superscriptaddress]{revtex4}
\pdfoutput=1 
\usepackage{amsmath,amssymb,eucal,graphicx}

\begin{document}
\title{Limiting shapes of Ising droplets, Ising fingers, and Ising solitons}
 
\author{P.~L.~Krapivsky}
\affiliation{Department of Physics, Boston University, Boston, MA 02215, USA}

\begin{abstract} 
We examine the evolution of an Ising ferromagnet endowed with zero-temperature single spin-flip dynamics. A large droplet of one phase in the sea of the opposite phase eventually disappears. An interesting behavior occurs in the intermediate regime when the droplet is still very large compared to the lattice spacing, but already very small compared to the initial size. In this regime the shape of the droplet is essentially deterministic (fluctuations are negligible in comparison with characteristic size). In two dimensions the shape is also universal, that is, independent on the initial shape. We analytically determine the limiting shape of the Ising droplet on the square lattice. When the initial state is a semi-infinite stripe of one phase in the sea of the opposite phase, it evolves into a finger which translates along its axis. We determine the limiting shape and the velocity of the Ising finger on the square lattice. An analog of the Ising finger on the cubic lattice is the translating Ising soliton. We show that far away from the tip, the cross-section of the Ising soliton coincides with the limiting shape of the two-dimensional Ising droplet and we determine a relation between the cross-section area, the distance from the tip, and the velocity of the soliton. 
\end{abstract}


\maketitle 

\section{Introduction}

At low temperatures, interfaces separating two broken-symmetry ordered phases generally shrink and eventually disappear.  This coarsening process is very complicated as it usually involves the evolution of numerous interfaces. The general understanding of coarsening (the emergence of the coarsening domain mosaic with a single characteristic scale growing in a universal manner) has been steadily improving over the last forty years \cite{Bray94,book}, although many concrete questions remain unanswered \cite{OKR}. Even in the simplest situation when the two-dimensional Ising ferromagnet endowed with a non-conservative dynamics is quenched from the temperature above the critical to zero temperature, very few analytical results have been established. (One such result describes the ultimate fate of the system, e.g. the probability to end up in a stripe state \cite{BKR09}; another \cite{LC07} gives the statistics of domain hulls in the case of the curvature-driven dynamics.) 

Here we shall examine the evolution of a single interface. We shall always assume that the Ising ferromagnet is subjected to zero-temperature non-conservative dynamics and we also consider the two-dimensional setting if not stated otherwise. Even in this situation the evolution of a single closed interface is not fully understood. (Equivalently, the evolution of a simply connected domain of the minority phase surrounded by the sea of the majority phase could not be generally `solved' \cite{Bray94,book}). The detailed evolution of a closed interface is not particularly interesting, but the asymptotic is as it is presumably universal (independent on the details of the initial condition). This is one of the goals of this study. 

To examine the evolution we need a precise description of the dynamics. Two most popular non-conservative zero-temperature dynamics are the single spin-flip dynamics (the chief example of the microscopic dynamics), and the time-dependent Ginzburg-Landau (TDGL) equation which is the prime  example of the macroscopic dynamics. A zero-temperature spin-flip dynamics forbids energy raising flips. Glauber's version \cite{glauber} of the single spin-flip dynamics specifically prescribes that energy conserving flips occur at a twice smaller rate than energy decreasing flips. For the Metropolis algorithm both energy conserving and energy decreasing flip rates are assumed to be equal. The energy-lowering spin-flip events do occur, yet their frequency is asymptotically negligible and therefore these subtle differences in the single spin-flip dynamics are irrelevant for the problems which we shall study. In the following, we set the rate of energy conserving flips to unity. 

Even at zero temperature the TDGL equation is difficult to analyze since mathematically it is a non-linear parabolic partial differential equation \cite{det}. Fortunately, in the interesting situation when the width of the interface is much smaller than its radius of curvature the TDGL equation reduces to a much simpler Lifshitz-Allen-Cahn (LAC) equation \cite{LAC} which asserts that the normal velocity of the interface is proportional to the local mean curvature.  (The terminology is not yet settled: The LAC equation is often termed the Allen-Cahn equation; sometimes the TDGL equation is called the Allen-Cahn equation; also in mathematical literature, people usually talk about mean-curvature flows \cite{curvature}.)

\begin{figure}
\hspace*{-0.27in}
\includegraphics[scale=0.87]{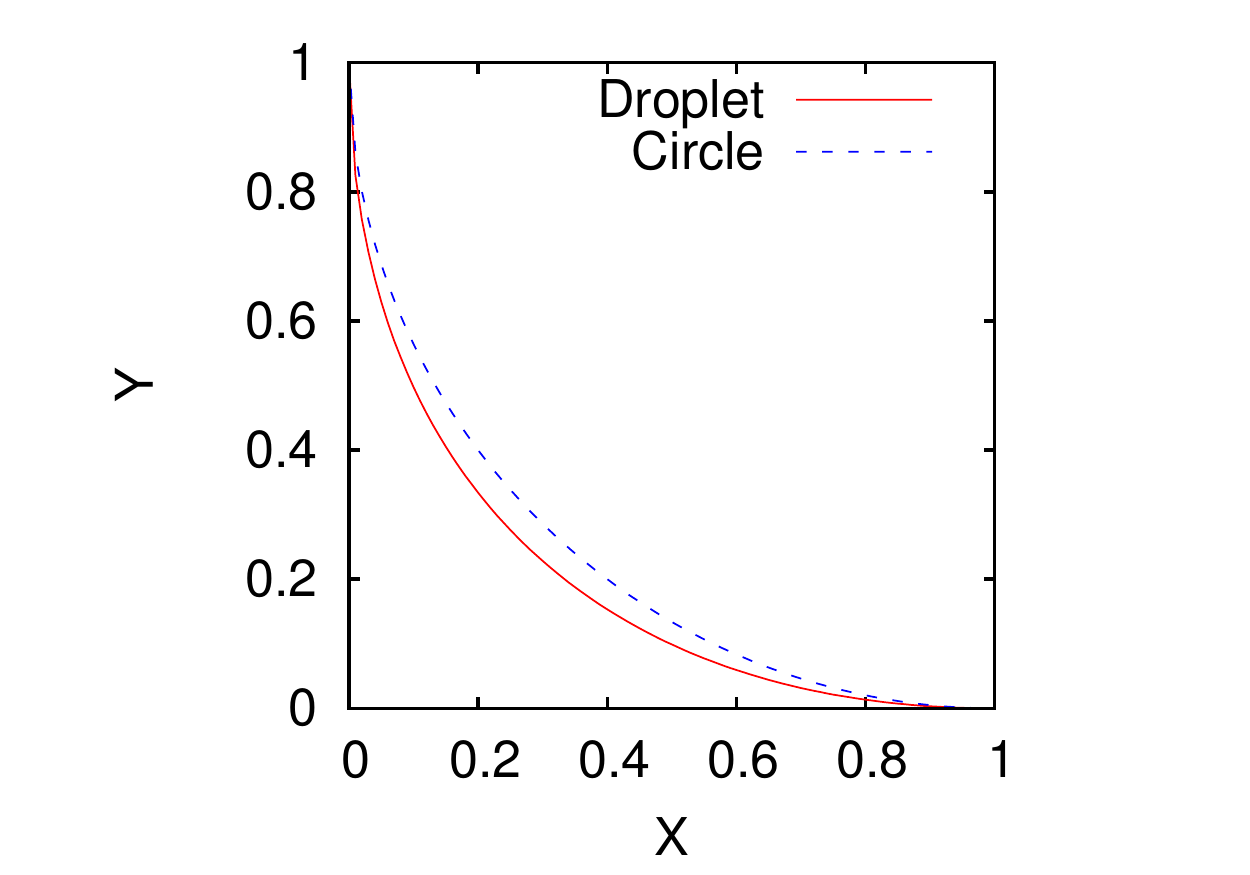}
\caption{(Color online) The quarter of the limiting shape of the Ising droplet and the quarter of the circle (the limiting shape corresponding to the macroscopic curvature-driven dynamics). The Ising droplet encloses the circle.} 
\label{d_shape}
\end{figure}

The asymptotic evolution of a single closed interface in two dimensions is fully understood in the realm of the LAC equation. Indeed, the Grayson theorem \cite{Grayson} asserts that the interface approaches to a circle, so at the final stage of shrinking to a point the interface is the circle. Thus with respect to the macroscopic non-conservative dynamics the limiting shape is the circle. What is the limiting shape with respect to the spin-flip dynamics? This question was previously investigated by Karma and Lobkovsky \cite{Alex} who exploited the self-similar behavior on the late stage of evolution and succeeded in reducing the problem to an ordinary differential equation which they solved numerically. In Sect.~\ref{I_droplet} we use an additional trick to reformulate the mathematical description in terms of the Stefan problem (more precisely, the diffusion equation with moving boundaries whose position is determined in the process of solution). This Stefan problem admits an exact self-similar solution which allows one to analytically determine the limiting shape of the Ising droplet. Figure \ref{d_shape} plots both limiting shapes. 

Infinite interfaces exhibit less universal evolution. Indeed, the spectrum of qualitatively different initial conditions is much more broad, and accordingly there are infinitely many different limiting shapes. For instance in the situation when the minority phase initially occupies the wedge (with opening angle smaller than $\pi$), the LAC equation admits a self-similar solution \cite{we} which gives the limiting shape (parametrized by the opening angle of the wedge). On the square lattice the most natural non-trivial `wedge' is the corner (the positive quadrant). The evolution of the corresponding infinite interface can be described in great details thanks to the mapping \cite{KD90,barma} on the symmetric exclusion process. Using this mapping together with results and methods developed in the studies of the symmetric exclusion process (see e.g.  \cite{DG09} and references therein) one can determine both the limiting shape and fluctuations \cite{KM}.  

Another interesting class of initial conditions corresponds to semi-infinite stripes. In this case, stripes quickly approach the limiting shape (known as the finger) which propagates with constant velocity along its axis. The shape of the finger corresponding to the macroscopic dynamics has been rediscovered a few times as it arises in numerous applications, e.g. in modeling of the motion of grain boundaries in an annealing piece of metal \cite{Mullins}, magneto-hydrodynamic models for the solar flares \cite{Low}, dendritic crystal growth \cite{other}, boundary renormalization group flows \cite{Z}, and various other problems in physics \cite{bakas} and mathematics \cite{curvature}. For spin-flip dynamics  (in this case we consider the finger which is parallel to one of the axes of the square lattice) the shape of the Ising finger was unknown. We compute the shape of the Ising finger in Sect.~\ref{I_finger}.

In Sect.~\ref{3d} we discuss challenges in computing the limiting shapes in three dimensions and look at a particular limiting shape corresponding to a translating Ising soliton. This is an approximately parabolic object which moves at a constant velocity along its axis (coinciding with the axis of the lattice). The final Sect.~\ref{conclusion} contains conclusions.

\section{Ising Droplet}
\label{I_droplet}

Any finite domain of one phase in the sea of the opposite phase disappears in a finite time that scales as the square of the characteristic size; in that sense, we qualitatively understand the shrinking of a finite domain. A more detailed {\em quantitative} understanding emerges in the long time limit since after a proper re-scaling, the interface admits a deterministic limiting shape as long as the droplet remains large. In this section we compute the limiting shape of the Ising droplet on the square lattice. 

The spin-flip dynamics is stochastic and this causes a number of subtle differences with the macroscopic dynamics (the deterministic LAC equation). To highlight one of the differences we note that the macroscopic dynamics always leads to the decrease of the area of the simply connected domain $\mathcal{D}$. To verify this assertion we write the LAC equation in the form $v_\text{normal}=-DK$, where $K$ is the  curvature and $D$ is the proportionality factor which has the dimension of the diffusion constant. The area $A(t)=\text{area}[\mathcal{D}(t)]$ then evolves according to
\begin{equation}
\label{A_curvature}
\frac{dA}{dt} = -\oint ds\,DK = -2\pi D
\end{equation}
The last step in Eq.~\eqref{A_curvature} is the consequence of the Gauss-Bonnet theorem. 

\begin{figure}
\centering
\includegraphics[scale=0.24]{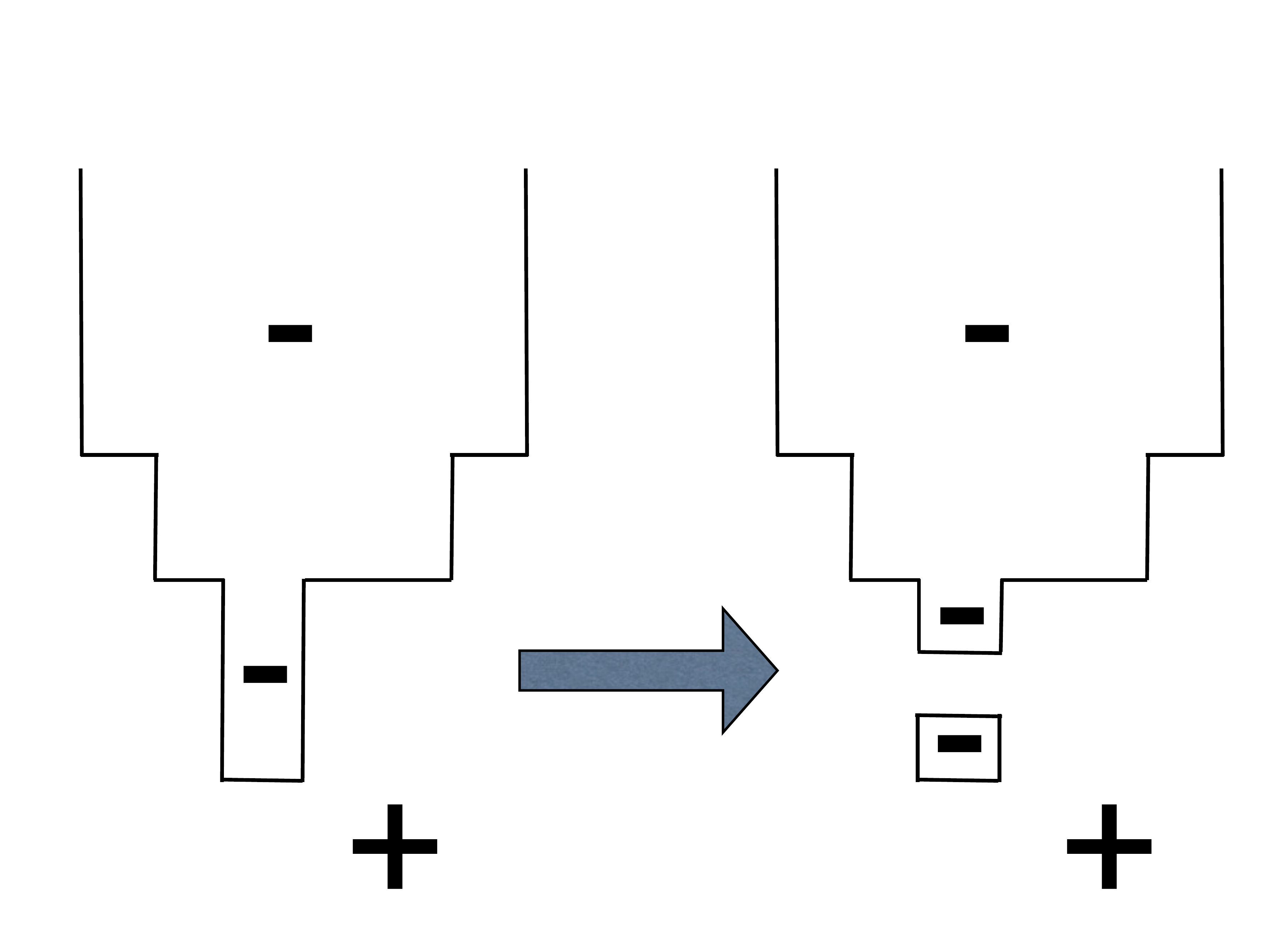}
\caption{An illustration of a rare emergence of a short-living tiny closed interface during the evolution of the finger.} 
\label{shedding}
\end{figure}

Equation \eqref{A_curvature} helps to understand coarsening dynamics of two-dimensional Ising \cite{LC07} and Potts \cite{LC10} systems, yet its deterministic nature disagrees with the microscopic spin-flip dynamics. For instance, the zero-temperature spin-flip dynamics allows area-raising moves. More precisely, any domain $\mathcal{D}$ can grow up to its rectangle envelope (which is defined to be the smallest rectangle aligned with the axes of the square lattice that contains $\mathcal{D}$). However, on average every domain shrinks. Another subtle feature of the stochastic spin-flip dynamics is that a single connected domain can evolve into a few disjoint domains (see Fig.~\ref{shedding}). Apart from pathological initial conditions, e.g. those which contain `tendrils' of width one (i.e. equal to the lattice spacing) or strips of width one, or rare separations of tiny drops (Fig.~\ref{shedding}), such break ups start to play a significant role only when the droplet becomes comparable with the lattice spacing; this late stage of evolution is clearly stochastic, but it is not interesting. With all these caveats we can talk about single domain, ignore fluctuations and focus on the limiting shape as long as the domain is very large compared to the lattice spacing. 

A few interesting mathematical papers discussed the replacement of the LAC equation with isotropic surface tension by an equation with anisotropic surface tension which is supposed to be a proper macroscopic equation corresponding to the spin-flip dynamics \cite{Spohn,law,Sow99,cerf}. We follow a similar approach, yet our goal is not to fully justify the governing equation (this is essentially achieved in previous papers), but rather to solve it analytically. 

The initial shape of the droplet is expected to become asymptotically irrelevant. For instance, the initial rectangular envelope $L_x\times L_y$ can be significantly different from the square, $L_x(t=0)\ne L_y(t=0)$, yet asymptotically the rectangular envelop of the droplet is approaching to the square. Denote by 
$2L(t)$ the size of this square at time $t$. For concreteness, let the envelope be the square $0\leq x,y\leq 2L$. Due to symmetry, we can limit ourselves to its quarter,  $0\leq x,y\leq L$; the boundary of the droplet thus goes from $(x,y)=(0,L)$ to $(x,y)=(L,0)$. This boundary can be represented by a staircase of kinks that can pile-up at the same site, but cannot pass through each other. It is more convenient to use a representation in terms of an exclusion process, that is a collection of particles which undergo a random walk and cannot occupy the same site. (Both these representations have appeared in the literature, see e.g. \cite{Rost,Liggett,spohn,KD90,barma,we,Alex}).  

\begin{figure}
\centering
\includegraphics[scale=0.36]{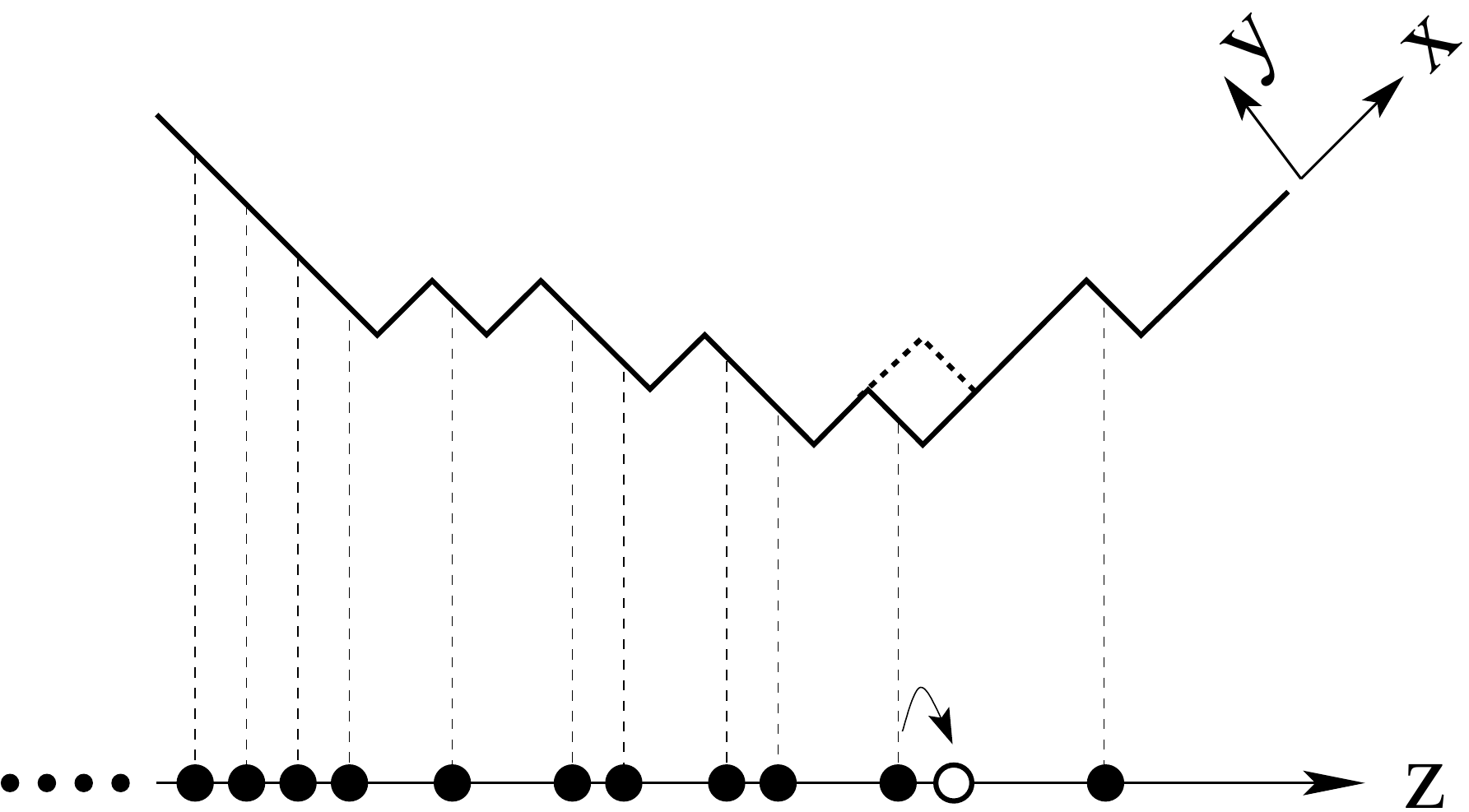}
\caption{An illustrative interface rotated by $\pi/4$ and the corresponding particle configuration. A spin-flip
event is shown together with the correspondence hop of the particle in the symmetric exclusion process.} 
\label{csp}
\end{figure}

The representation in terms of the symmetric exclusion process becomes evident after rotating counter-clockwise by angle $\pi/4$ around the origin and projecting the boundary onto the horizontal line (see Fig.~\ref{csp}). We put a particle on the bond (leave the bond empty) if the corresponding bond on the interface goes along co-diagonal (diagonal). The particles can occupy the lattice sites, with no more than one particle per site, and the particles undergo the symmetric exclusion process. 

In the long time `hydrodynamic' limit we employ a continuum description. The particle density $n(z,t)$ satisfies a diffusion equation
\begin{equation}
\label{diff}
\frac{\partial n}{\partial t} = \frac{\partial^2 n}{\partial z^2}
\end{equation}
on a shrinking interval $-L(t)\leq z\leq L(t)$.
The boundary conditions are
\begin{equation}
\label{bound}
n(-L(t),t)=1, \qquad n(L(t),t)=0
\end{equation}
It is sufficient to seek a self-similar solution which depends on $z$ and $t$ only through a combination $z/L(t)$, 
\begin{equation}
\label{scaling}
n(z,t)=N(Z), \quad Z=z/L(t)
\end{equation}
Plugging \eqref{scaling} into \eqref{diff} we obtain
\begin{equation}
\label{diff-ODE}
-L \dot L Z N' = N''
\end{equation}
where prime (dot) denotes differentiation with respect to $Z$ (time $t$). We now notice that the width $L(t)$ shrinks at a rate that is equal to the flux of particles:
\begin{equation*}
\dot L = \frac{\partial n}{\partial z}\Big|_{z=L(t)} = \frac{1}{L}\,N'(1)
\end{equation*}
Introducing the parameter $b$ defined via
\begin{equation}
\label{b_def}
L\dot L = -2b = N'(1)
\end{equation}
we re-write \eqref{diff-ODE} as
\begin{equation}
\label{ODE}
N'' = 2bZ N'
\end{equation}
Integrating \eqref{ODE} subject to $N'(1)=-2b$ we obtain
\begin{equation}
\label{n'-sol}
N' = -2b\,e^{bZ^2-b}
\end{equation}
Integrating Eq.~\eqref{n'-sol} and using $N(1)=0$ we get
\begin{equation}
\label{n-sol}
N(Z) = 2b\int_Z^1 dv\,e^{bv^2-b}
\end{equation}
The boundary condition $N(-1)=1$, or equivalently $N(0)=1/2$, yields
\begin{equation}
\label{b}
1=4b\int_0^1 dv\,e^{bv^2-b}
\end{equation}
from which $b \approx 0.3051025211$. In terms of original variables, the interface is implicitly given by 
\begin{equation}
\label{yx}
y(x,t)=\int_{x-y}^\infty dz\,n(z,t)
\end{equation}
Writing
\begin{equation}
X=\frac{x}{L(t)}\,,\quad Y=\frac{y}{L(t)}
\end{equation}
and using \eqref{n-sol} we re-write \eqref{yx} as
\begin{equation}
\label{interface}
Y=1-e^{bu^2-b}-2bu\int_u^1 dv\,e^{bv^2-b}\,, \quad u\equiv X-Y
\end{equation}

The limiting shape \eqref{interface} together with the circle (the limiting shape corresponding to the LAC equation) are plotted on Fig.~\ref{d_shape}. The Ising droplet encloses the circle when both limiting shapes are re-scaled in such a way that their rectangular envelops are identical $[0,2]\times [0,2]$ squares. For instance, the diagonal point on the Ising droplet lies at  
\begin{equation*}
X_*=Y_*=1-e^{-b} = 0.262952192433887454\ldots
\end{equation*}
while for the circle
\begin{equation*}
X_*=Y_*=1-\frac{1}{\sqrt{2}} = 0.292893218813452427\ldots
\end{equation*}
Near the axes the Ising droplet is more flat, e.g. 
\begin{equation*}
Y=b\,(1-X)^2+\ldots
\end{equation*}
when $X\to 1$, while in the case of the circle
\begin{equation*}
Y=\frac{1}{2}\,(1-X)^2+\ldots
\end{equation*}

Chayes, Schonmann and Swindle \cite{law} proved that the area of the droplet decreases with average rate 4. This theorem provides a useful consistency check. Let us first compute the area under a quarter of the interface 
\begin{equation}
\label{area}
A=L^2\int_0^1dX\int_0^1 dY
\end{equation}
Changing variables, $(X,Y)\to (u=X-Y,Y)$, and noting that the Jacobian is equal to unity, $\frac{D(u,Y)}{D(X,Y)}=1$, we get
\begin{equation}
\label{int}
A=2L^2\int_0^1 du\,Y(u)
\end{equation}
Equation \eqref{interface} gives an explicit expression for $Y=Y(u)$. Plugging it into \eqref{int} we compute the integral and find that the area of the droplet $\mathcal{A}=4L^2-4A$ is given by
\begin{equation}
\mathcal{A}=4L^2\left[2\int_0^1 du\,(1+bu^2)e^{bu^2-b} -1\right]
\end{equation}
Differentiating and using $L\dot L=-2b$ we get \cite{constants}
\begin{equation}
\label{A_flip}
\dot{\mathcal{A}} =-16b\left[2\int_0^1 du\,(1+bu^2)e^{bu^2-b} -1\right]=-4
\end{equation}
The last relation is established by using \eqref{b} and 
\begin{equation}
\label{b2}
2\int_0^1 du\,e^{bu^2-b}\,bu^2 = 1- (4b)^{-1}
\end{equation}
which is derived from  \eqref{b} through integration by part. 

\section{Ising Finger}
\label{I_finger}

\begin{figure}
\hspace*{-0.1in}
\includegraphics[scale=0.45]{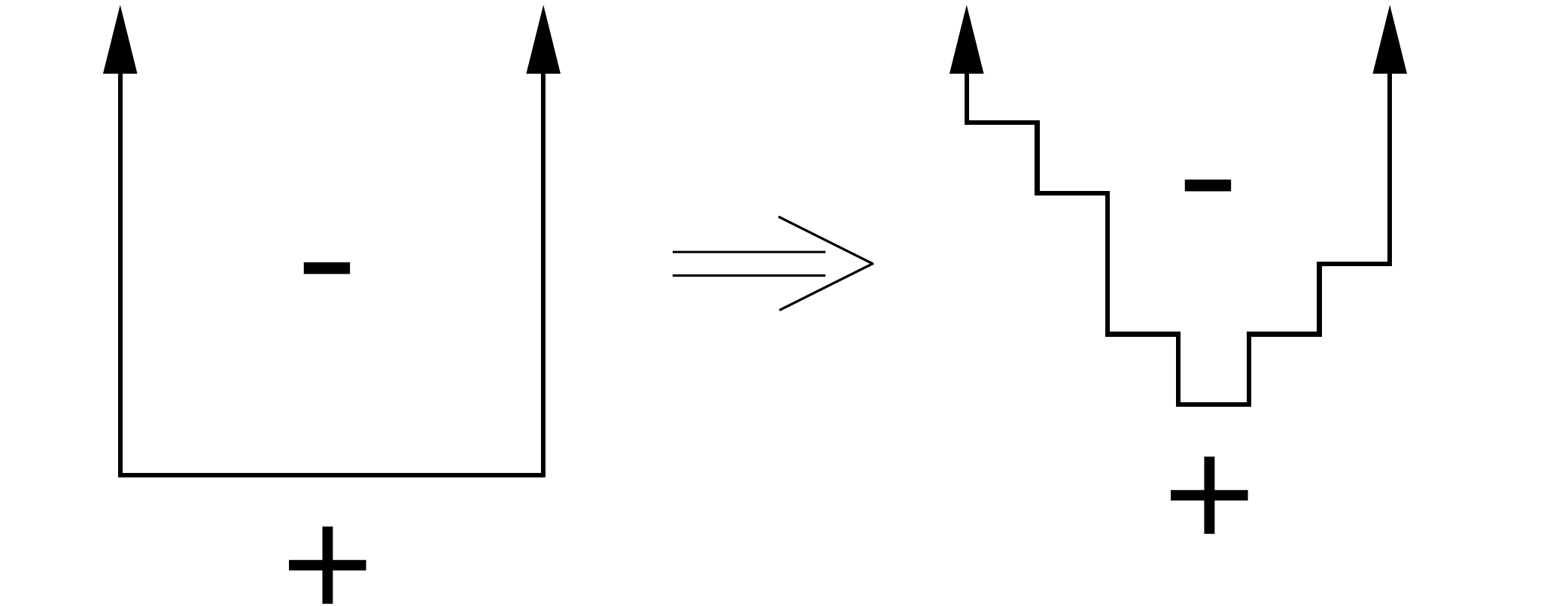}
\caption{Schematic illustration of the evolution of a semi-infinite strip (a rectangular finger of the minority minus phase surrounded by the majority plus phase). On the right-hand side, the flip of the lowest minority spin, the tip spin in this example, is an irreversible process that causes the minimum height of the finger to advance by one.} 
\label{finger_ill}
\end{figure}

Here we consider the finger geometry (Fig.~\ref{finger_ill}), e.g. we assume that the minority phase initially occupies the semi-infinite region $y>0$ and $|x|<L$.  The interesting regime is $t\gg L^2$, where the two corners of the initial finger interact and the finger relaxes to a limiting shape that eventually recedes at constant velocity.  In a reference frame moving with the finger, the interface $y(x)$ is thus stationary.

We consider the spin-flip dynamics which is stochastic, so the area of the finger can occasionally increase, although on average it decreases. The rectangular envelop is now a semi-infinite region $y>h$ and $|x|<L$ where $h(t)$ is the current height of the tip. More precisely, the tip is formed by all adjacent spins on the lowest height.  If the tip of the finger contains a single spin (as on the right-hand side of Fig.~\ref{finger_ill}), then when this spin flips, the fingertip irreversibly advances by one unit.  The finger can shed disconnected pieces whenever the tip of the finger has the width equal to one and the height greater than one (see Fig.~\ref{shedding}).  Here we consider a very wide finger, $L\gg 1$, and in this situation the above subtleties are asymptotically negligible, so we ignore fluctuations and shedding events and focus on the limiting shape. (For narrow fingers the stochastic effects are important. The full description of the evolution of the finger of the least possible width 2 is highly non-trivial and unknown.)

\begin{figure}
\centering
\includegraphics[scale=0.7]{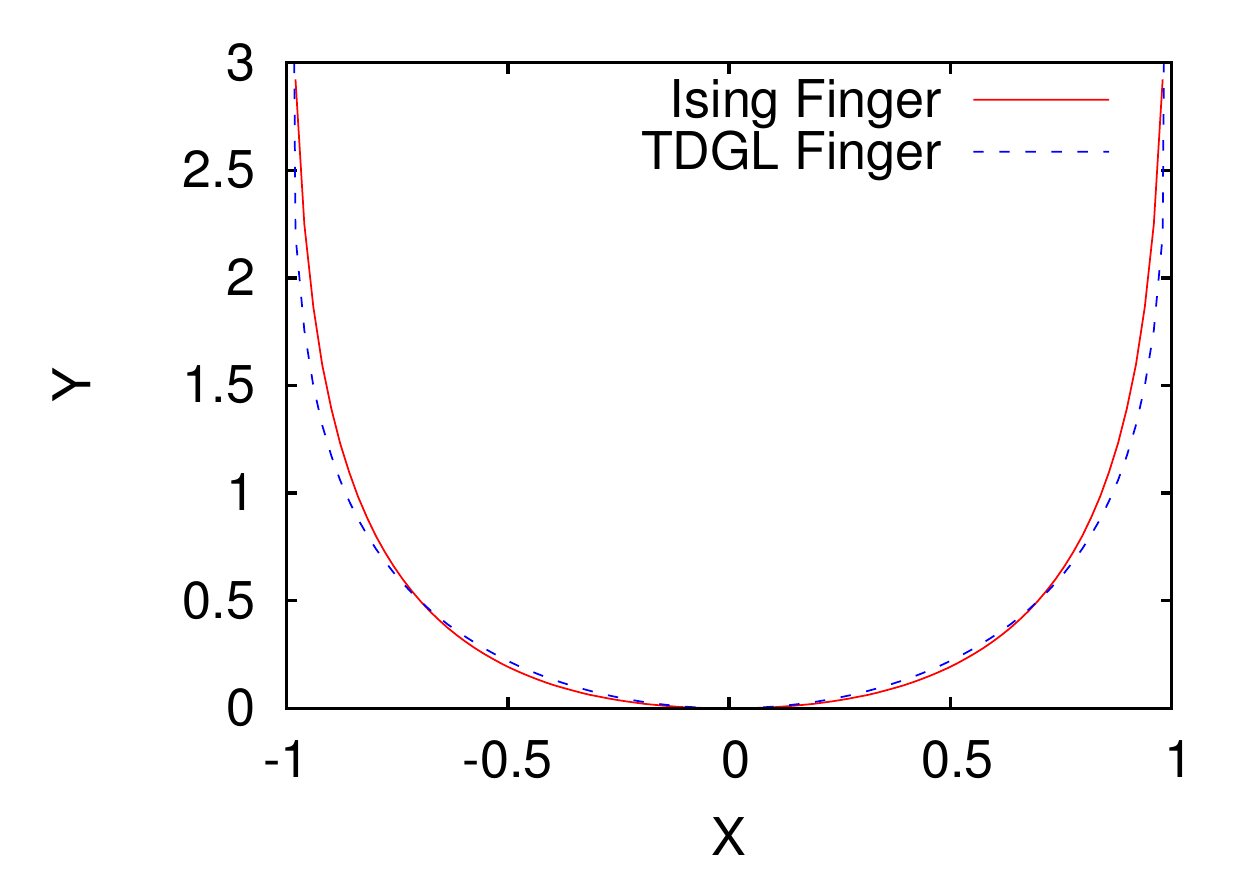}
\caption{(Color online) The limiting shape of the Ising finger, Eq.~\eqref{finger-ising}, corresponding to the microscopic spin-flip dynamics; the limiting shape of the TDGL finger,  Eq.~\eqref{finger}, corresponding to the macroscopic curvature-driven dynamics.} 
\label{f_shape}
\end{figure}

Due to symmetry we can limit ourselves to the region $0<x<L$ and $y>0$. The governing equation for the interface shape $y(x,t)$ is \cite{Alex}
\begin{equation}
\label{2d}
y_{t}=\frac{y_{xx}}{(1+y_x)^2}
\end{equation}
where $y_t=\frac{\partial y}{\partial t}, ~y_x=\frac{\partial y}{\partial x}$, etc. In an upward moving reference system in which the surface is stationary
\begin{equation}
v=\frac{y_{xx}}{(1+y_x)^2}
\end{equation}
This equation must have a solution satisfying the boundary conditions $y(0)=0$ and $y(L)=\infty$. These conditions fix the velocity
\begin{equation}
\label{finger-velocity}
v=\frac{1}{L}
\end{equation}
The shape of the finger is 
\begin{equation}
\label{finger-ising}
Y = -\ln(1-X)-X\,,\quad (X,Y)=\left(\frac{x}{L},\frac{y}{L}\right)
\end{equation}
This theoretical prediction perfectly agrees with previous simulation results (cf. Fig.~\ref{f_shape} with Fig.~8 of \cite{we}). 

It is interesting to compare the Ising finger \eqref{finger-ising} with the finger that arises if the system evolves according to the TDGL equation, that is the mean-curvature evolution (Fig.~\ref{f_shape}). In the latter framework, the interface $y(x)$ satisfies the equation $y_{xx}=v(1+y_x^2)$ in the upward moving reference system in which the surface is stationary.  Integrating this equation and imposing the boundary condition $y\to\infty$ when $|x|\to L$, one arrives at the following TDGL finger 
\begin{equation}
\label{finger}
Y = -\frac{2}{\pi}\,\ln\left[\cos\left(\frac{\pi X}{2}\right)\right]
\end{equation}
This finger recedes at a constant velocity that is given by $v=y''(x=0)=\pi/2L$.  The expression \eqref{finger} for the finger shape has been derived by Mullins in the context of modeling the motion of grain boundaries \cite{Mullins}, and it has subsequently arisen in numerous applications \cite{Low,other,Z,we,bakas,curvature}; this finger also appears under the name {\em grim-reaper} in the mathematics literature and {\em hair-pin} \cite{Z} in the physics literature.

\section{Ising Soliton} 
\label{3d} 

In three dimensions, even in the realm of the LAC equation (equivalently the mean-curvature evolution)  little is known about the limiting shapes. If the initial domain is convex and compact, it approaches to a ball and eventually shrinks to a point. Thus in this setting the limiting shape is a sphere \cite{H84} and the behavior is similar to the behavior observed in two dimensions \cite{Grayson}. However, in three dimensions the assumption of convexity is important --- a compact closed interface which is topologically a sphere may become singular before it shrinks to a point. The simplest example is provided by a dumbbell with sufficiently thin neck. The classification of possible limiting shapes (the surfaces of the shrinking LAC droplets) and the description of singularities of closed interfaces undergoing the mean-curvature evolution is an active research area, see e.g. Refs.~\cite{ES91,AAG95,OS}. 

In the case of zero-temperature spin-flip dynamics, not a single limiting shape is known in three dimensions. Indeed, even an equation governing the evolution of the interface in three dimensions is unknown. In the simplest case when when the initial domain is convex, or almost convex, the limiting shape should be closed to the sphere. Simulations \cite{Alex} confirm this expectation.

Consider now possible analogs of the two-dimensional finger. Let us first discuss the mean-curvature evolution. In three dimensions there are various possibilities, e.g. one can start with a semi-infinite bar, or consider rotationally invariant initial conditions. The former case is more similar to the two-dimensional finger, but mathematically the problem is challenging since even in the reference frame moving with the finger one must solve a non-linear partial differential equation. In the rotationally invariant setting the problem reduces to an ordinary differential equation. The limiting shape is very different from the finger, viz. the width of the cross-section diverges with the distance from the tip. This solution is usually termed the (rotationally invariant) translating soliton and it has been investigated in Refs.~\cite{AW94,CSS07,OS}. Using the cylindrical coordinates and representing the interface in the form $r=r(z,t)$ we reduce the mean-curvature evolution equation to
\begin{equation}
\label{soliton:curv}
r_t = \frac{r_{zz}}{1+r_z^2} - \frac{1}{r}
\end{equation}
This equation admits a family of solutions parametrized by the soliton speed $v>0$. 
For the soliton moving with constant speed $v$ along the $z-$direction we write
\begin{equation}
\label{Z_def}
r(z,t) = v^{-1} R(Z), \quad z-vt = v^{-1}Z 
\end{equation}
and recast Eq.~\eqref{soliton:curv} into an ordinary differential equation
\begin{equation}
\label{soliton:GE}
-R' = \frac{R''}{1+(R')^2} - \frac{1}{R}
\end{equation}
where prime denotes the derivative with respect to $Z$. Equation \eqref{soliton:GE} cannot be solved in quadratures, but one can readily extract asymptotic behaviors near the tip of the soliton (without loss of generality we set the position of the tip to $Z=0$) and far away from the tip. Near the tip, $Z\to +0$, one finds that $R^2$ has a regular asymptotic expansion in powers of $Z$: 
\begin{equation}
\label{soliton:near}
R^2 = 4Z - \tfrac{1}{2}Z^2 - \tfrac{5}{72}Z^3 + \ldots
\end{equation}
Far away from the tip, $Z\to\infty$, the expansion is more cumbersome as it contains logarithms:
\begin{equation}
\label{soliton:far}
R^2 = 2Z + \ln Z + \ldots
\end{equation}
Overall, the interface is approximately parabolic. 

Consider now an Ising soliton on the cubic lattice which moves with a constant velocity along $z$ axis. This Ising soliton is an analog of the rotationally invariant soliton \eqref{soliton:GE}--\eqref{soliton:far}. We cannot provide the detailed description of the Ising soliton as we don't even know the governing evolution equation. Far away from the tip, however, the problem is essentially two-dimensional. (The same is valid in the case of the macroscopic dynamics --- far away from the tip the LAC equation \eqref{soliton:curv} reduces to the two-dimensional LAC equation $r_t=-1/r$.) Therefore the cross-section of the Ising soliton must be identical to the Ising droplet which was studied in Sect.~\ref{I_droplet}. Let $[-L,L]\times [-L,L]$, where $L=L(z,t)$, be the rectangular envelop of the Ising soliton. The soliton moves without changing its shape, so $L(z,t)=L(Z)$ with $Z$ defined in \eqref{Z_def}. Plugging $L(Z)$ into \eqref{b_def} we get
\begin{equation*}
-v^2 L \frac{dL}{dZ} = -2b
\end{equation*}
from which we find the leading behavior far away from the tip:
\begin{equation}
\label{soliton:far_Ising}
L^2 = \frac{4b}{v^2}\,Z
\end{equation}
This behavior is similar to \eqref{soliton:far} describing the rotationally invariant soliton once we notice that $vL$ plays the role of $R$, cf. \eqref{Z_def}. Thus the shape of the Ising soliton is asymptotically parabolic, e.g. the area of the cross-section scales linearly with the distance from the tip in the $Z\to\infty$ limit; the cross-section is not a disk even far away from the tip, it is actually asymptotically identical to the Ising droplet on the square lattice. 

To determine the entire shape of the Ising soliton one must know an equation governing the evolution of the interface in three dimensions and even then one would rely on numerical integration since even in the reference frame moving with the soliton a governing equation would turn into a non-linear partial differential equation. One possible direction is to {\em guess} a governing equation and then try to determine the behavior near the tip where one can employ asymptotic methods. An attempt of such an analysis is given in Appendix~\ref{app}. 

\section{Conclusions}
\label{conclusion}

We have performed an analytical study of the limiting shapes arising in the context of the zero-temperature single spin-flip dynamics. We have determined the limiting shape of the shrinking Ising droplet and the limiting shape of the Ising finger which moves along its axis (coinciding with one of the two axes of the square lattice). We have investigated the limiting shapes only on the square lattice; an analytical computation of the limiting shapes on the hexagonal lattice appears feasible and it would be interesting to perform such a calculation. 

There is much more room for diverse limiting shapes on the cubic lattice. Nothing is known about these limiting shapes. The chief reason is the lack of the equation describing the evolution of interfaces on the cubic lattice. There are also intrinsic mathematical difficulties in analyzing such would-be governing equations. For one particular limiting shape, the translating Ising soliton which moves along its axis (coinciding with one of the three axes of the cubic lattice), one can circumvent the aforementioned challenges and extract partial analytic information. We have shown that asymptotically (that is, far away from the tip) the surface of the translating Ising soliton is parabolic, e.g. the cross-section area grows linearly with the distance from the tip. Furthermore, the cross-section of the Ising soliton coincides with the limiting shape of the Ising droplet on the square lattice. Making a plausible guess about the governing evolution equation, we have deduced the shape of the translating Ising soliton near its tip (Appendix~\ref{app}). 

\acknowledgments
I have benefitted from pleasant and stimulating interactions with S. N. Majumdar, K. Mallick, J. Olejarz, S. Redner, J.~Tailleur, and D. Volovik; I am additionally grateful to D. Volovik for assistance. 

\appendix
\section{Tip of the Translating Ising Soliton}
\label{app}

We need to know the equation governing the evolution of the interface on the cubic lattice. In a related (but simpler) problem of spin-flip dynamics in a magnetic field, the governing evolution equation for the interface has been guessed in recent study \cite{JPSK}. This guess has been guided by symmetry considerations (the governing equation for $z(x,y;t)$ should be invariant under the change of any coordinate pair) and by the requirement that it should reduce to the two-dimensional evolution equation in appropriate settings. Two functionally independent simple equations have been found \cite{JPSK}, as well as families of equations built from the independent solutions. Only the two independent equations seemed sufficiently simple and plausible, one of the two simple equations was found to be in excellent agreement with simulations. A similar program is possible in the present case, and one gets
the analog of equation which is expected to be exact is
\begin{equation}
\label{Ising_3d}
z_t = \frac{\big(1+\frac{1}{z_x+z_y}\big)^2}{\big(1+\frac{1}{z_x}\big)^2\big(1+\frac{1}{z_y}\big)^2} 
\left[\frac{z_{xx}}{z_x^2} - \frac{z_{xy}}{z_xz_y} +  \frac{z_{yy}}{z_y^2}\right]
\end{equation}
This equation is a generalization of Eq.~\eqref{2d} governing the evolution of the interfaces on the square lattice and it is an analog of the equation describing  the evolution of the interfaces on the cubic lattice in the presence of the magnetic field \cite{JPSK}.  

Equation \eqref{Ising_3d} applies to the region $x>0, y>0$ where $z_x>0, z_y>0$. Generally the first derivatives should be written as $|z_x|$ and  $|z_y|$ and then the equation will be applicable everywhere. Due to symmetry we can limit ourselves to the region $x>0, y>0$, so the form \eqref{Ising_3d} suffices. Equation \eqref{Ising_3d} is harder to judge than equations analyzed in  \cite{JPSK} as we cannot solve \eqref{Ising_3d} analytically. Postponing a careful consideration of the agreement between the predictions of  \eqref{Ising_3d} and numerical results for the future, let's just examine the behavior near the tip. First, we notice that for the Ising soliton which moves with constant velocity $v$ along $z$ axis Eq.~\eqref{Ising_3d} 
reduces to
\begin{equation}
\label{Ising_sol_eq}
v = \frac{\big(1+\frac{1}{z_x+z_y}\big)^2}{\big(1+\frac{1}{z_x}\big)^2\big(1+\frac{1}{z_y}\big)^2} 
\left[\frac{z_{xx}}{z_x^2} - \frac{z_{xy}}{z_xz_y} +  \frac{z_{yy}}{z_y^2}\right]
\end{equation}
An additional transformation 
\begin{equation}
\label{xyz}
x= v^{-1}X, \quad y= v^{-1}Y, \quad z-vt= v^{-1}Z
\end{equation}
allows us to recast Eq.~\eqref{Ising_sol_eq} to $v-$independent form
\begin{equation}
\label{Ising_soliton_eq}
\frac{\big(1+\frac{1}{Z_X}\big)^2\big(1+\frac{1}{Z_Y}\big)^2}{\big(1+\frac{1}{Z_X+Z_Y}\big)^2} = 
\frac{Z_{XX}}{Z_X^2} - \frac{Z_{XY}}{Z_X Z_Y} +  \frac{Z_{YY}}{Z_Y^2}\
\end{equation}
Near the tip, that is, when $0<X\ll 1$ and $0<Y\ll 1$, we seek the solution as an expansion in the form analogous to \eqref{soliton:near}, namely
\begin{equation}
\label{soliton:near_I}
Z = C(X^2+2\lambda XY + Y^2) + \ldots
\end{equation}
Plugging \eqref{soliton:near_I} into \eqref{Ising_soliton_eq} and keeping only the leading terms we find the consistency when an `asymmetry' parameter $\lambda$ is the root of the cubic equation 
$\lambda^3-3\lambda+2=0$ and the amplitude is given by $C=\frac{1}{2}\,(1+\lambda)^2$. Since $\lambda^3-3\lambda+2 = (\lambda-1)^2(\lambda+2)$ there are two roots, $\lambda=1$ and $\lambda=-2$. The second root is inappropriate since the resulting quadratic form $X^2-4XY+Y^2$ vanishes in the $X>0, Y>0$. Hence $\lambda=1$ and $C=2$, so near the tip
 \begin{equation}
\label{soliton:near_tip}
Z = 2(|X|+ |Y|)^2 + \ldots
\end{equation}
where we have written the solution in the form which is valid for all $X, Y$ satisfying $|X|\ll 1$ and $|Y|\ll 1$. More precisely, \eqref{soliton:near_tip} is valid when
\begin{equation}
\label{bounds}
v\ll |X|\ll 1, \quad v\ll |Y|\ll 1
\end{equation}
Indeed, the continuum description underlying the usage of evolution equations like \eqref{Ising_3d} applies only when the distance from the tip far exceeds the lattice spacing. In other words, $|x|\gg 1$ and $|y|\gg 1$, and this in conjunction with \eqref{xyz} gives the lower bounds in \eqref{bounds}. Not surprisingly, the velocity must be very small. The same applies, of course, to the two-dimensional finger, viz. the continuum description leading to the limiting shape \eqref{finger-ising} is valid when the corresponding velocity \eqref{finger-velocity} is very small.


\begin{thebibliography}{99}

\bibitem{Bray94} 
    A.~J.~Bray, Adv.\ Phys.\ {\bf 43}, 357 (1994).

\bibitem{book}   
    P. L. Krapivsky, S. Redner and E. Ben-Naim,  {\it  A
    Kinetic View of Statistical Physics} (Cambridge: Cambridge University Press, 2010).

\bibitem{OKR}
   J. Olejarz,  P.~L.~Krapivsky, and S.~Redner, Phys.\ Rev.\ E {\bf 83}, 030104 (2011);
   Phys.\ Rev.\ E {\bf 83}, 051104 (2011). 

\bibitem{BKR09} 
    K. Barros, P. L. Krapivsky, and S. Redner, Phys.\ Rev.\ E {\bf 80}, 040101 (2009).

\bibitem{LC07} 
     J. J. Arenzon, A. J. Bray, L. F. Cugliandolo, and A. Sicilia, Phys. Rev. Lett. {\bf 98}, 145701 (2007); 
    A. Sicilia, J. J. Arenzon, A. J. Bray, and L. F. Cugliandolo, Phys. Rev. E {\bf 76}, 061116 (2007).

\bibitem{glauber} 
    R.~J.~Glauber, J. Math.\ Phys.\ {\bf 4}, 294 (1963).

\bibitem{det} 
    At zero temperature, the TDGL equation is a deterministic partial differential equation; 
    at a positive temperature, the TDGL equation 
    is a stochastic partial differential equation. Microscopic spin-flip dynamics remain 
    stochastic even at $T=0$.
  
\bibitem{LAC}
    I. M. Lifshitz, Zh. Eksp. Teor. Fiz. {\bf 42}, 1354 (1962) [Sov. Phys. JETP {\bf 15}, 939 (1962)];
    S.~M.~Allen and J.~W.~Cahn, Acta Metall. {\bf 27}, 1085 (1979).

\bibitem{curvature}
    X.-P. Zhu, {\it Lectures on Mean Curvature Flows} (International Press, Somerville, 2002); 
    C. Mantegazza, {\it Lecture Notes on Mean Curvature Flow} (Birkh\"auser Verlag AG, Basel, 2011). 

\bibitem{Grayson}
     M. A. Grayson, J. Differ. Geom. {\bf 26}, 285 (1987); Duke Math. J. {\bf 58}, 555 (1989). 

\bibitem{Alex}
    A.~Karma and A.~E.~Lobkovsky, Phys.\ Rev.\ E {\bf 71}, 036114 (2005).

\bibitem{we}
    P.~L.~Krapivsky, S.~Redner, and J.~Tailleur, Phys.\ Rev.\ E {\bf 69}, 026125 (2004).
   
\bibitem{KD90}
   D.~Kandel and E.~Domany, J.\ Stat.\ Phys. {\bf 58}, 685 (1990). 
    
\bibitem{barma}
   M.~Barma, J.\ Phys.\ A {\bf 25}, L693 (1992). 
    
\bibitem{DG09} 
    B. Derrida and A. Gerschenfeld, J. Stat. Phys. {\bf 136}, 1 (2009). 

\bibitem{KM}
   P.~L.~Krapivsky and K. Mallick, in preparation. 
   
\bibitem{Mullins}
    W.~W.~Mullins, J.\ Appl.\ Phys.\ {\bf 27}, 900 (1956); {\bf 28}, 333 (1957). 

\bibitem{Low}
    B.~C.~Low, Astrophys.\ J. {\bf 181}, 209 (1973); {\bf 184}, 917 (1973). 

\bibitem{other}
    T.~Ohta, D.~Jasnow, and K.~Kawasaki, Phys.\ Rev.\ Lett. {\bf 49}, 1223 (1982); 
    R. C. Brower, D. A. Kessler, J. Koplik and H. Levine, Phys.\ Rev.\  A {\bf 29}, 1335 (1984). 

\bibitem{Z}
    S. L. Lukyanov, E. S. Vitchev, and A. B. Zamolodchikov, Nucl. Phys. B {\bf 683}, 
    423 (2004); S. L. Lukyanov, A. M. Tsvelik, and A. B. Zamolodchikov,
    Nucl. Phys. B {\bf 719}, 103 (2005); S. L. Lukyanov, and A. B. Zamolodchikov,
    Nucl. Phys. B {\bf 744}, 295 (2006).

\bibitem{bakas}
    I. Bakas and C. Sourdis, J. High Energy Phys. {\bf JHEP} 0706 (2007) 057. 

\bibitem{LC10} 
    M. P. O. Loureiro, J. J. Arenzon, L. F. Cugliandolo, and A. Sicilia, Phys. Rev. E {\bf 81}, 021129 (2010). 

\bibitem{Spohn}  
    H. Spohn, J. Stat. Phys. {\bf 79}, 1081 (1993).

\bibitem{law}  
    L. Chayes, R. H. Schonmann, and G. Swindle, J. Stat. Phys. {\bf 79}, 821 (1995).

\bibitem{Sow99}  
    R. B. Sowers, J. Funct. Anal. {\bf 169}, 421 (1999).

\bibitem{cerf}  
    R. Cerf and S. Louhichi, Probab. Theory Relat. Fields {\bf 137}, 379 (2007).  

\bibitem{Rost}
    H.~Rost, Z. Wahrsch. Verw. Gebiete {\bf 58}, 41 (1981). 

\bibitem{Liggett}
    T.~M.~Liggett, {\it Interacting Particle Systems} (Springer, New York, 1985). 

\bibitem{spohn} 
    H. Spohn, {\it Large Scale Dynamics of Interacting Particles} (Springer, Berlin, 1991).

\bibitem{constants} 
    The area law \eqref{A_flip} for the spin-flip dynamics is dimensionless and it reflects 
    that we have chosen the lattice spacing 
    as the unit of time and the inverse flip rate of energy conserving flips as the unit of time. 
    The area law \eqref{A_curvature} for the curvature driven dynamics is written in the dimensionful form.   

\bibitem{H84}
     G. Huisken, J. Differ. Geom. {\bf 20}, 237 (1984). 

\bibitem{ES91}
     L. Evans  and J. Spruck, J. Differ. Geom. {\bf 33}, 635 (1991); J. Geom. Anal. {\bf 2}, 121 (1992). 
     
\bibitem{AAG95}     
     S. Altschuler, S. Angenent, and Y. Giga, J. Geom. Anal. {\bf 5}, 293 (1995). 

\bibitem{OS}     
     K. Olsen and C. Sourdis, J. Phys. A {\bf 42}, 355205 (2009). 
     
\bibitem{AW94}     
     S. Altschuler and L. Wu, Calc. Var. {\bf 2}, 101 (1994). 
     
\bibitem{CSS07}     
     J. Clutterbuck, O. Schnurer, and F. Schulze, Calc. Var. {\bf 29}, 281 (2007). 
     
\bibitem{JPSK}
      J. Olejarz, P.~L.~Krapivsky, S. Redner, and K. Mallick, Phys.\ Rev.\ Lett. {\bf 108}, 016102 (2012). 

\end{thebibliography}
\end{document}